\documentclass[aps,showpacs,pra]{revtex4}

\usepackage[tbtags]{amsmath}
\usepackage{amssymb}
\usepackage{amsfonts}
\usepackage[dvips]{graphicx,color}

\begin{document}
\title{Analysis of the time-energy entanglement of down-converted photon pairs by correlated single photon interference}
\author{Changliang Ren}
\author{Holger F. Hofmann}
\email{hofmann@hiroshima-u.ac.jp} \affiliation{ Graduate School of
Advanced Sciences of Matter, Hiroshima University, Kagamiyama 1-3-1,
Higashi Hiroshima 739-8530, Japan} \affiliation{JST, CREST,
Sanbancho 5, Chiyoda-ku, Tokyo 102-0075, Japan}

\begin{abstract}
The time-energy entanglement of down-converted photon pairs is
particularly difficult to characterize because direct measurements
of photon arrival times are limited by the temporal resolution of
photon detection. Here, we explore an alternative possibility of
characterizing the temporal coherence of two-photon wavefunctions
using single photon interference with weak coherent light.
Specifically, this method makes use of the fact that down-conversion
generates a coherent superposition of vacuum and two-photon states,
so that the coincidence count rates for photon pairs after
interference with weak coherent light are given by a superposition
of the two-photon wavefunction from the down-conversion with the
product wavefunction defined by the weak coherent references. By
observing the frequency dependent interference pattern, it is
possible to reconstruct the amplitudes and the phases of the
two-photon wavefunction within the bandwidth of the reference pulses
used in the single photon interferences. The correlated single
photon interferences therefore provide a direct map of the entangled
two-photon wavefunction generated in the down-conversion process.
\end{abstract}


\maketitle

\section{Introduction}

Photon pairs generated by spontaneous parametric
down-conversion are usually entangled in time and energy
\cite{Franson}, and a wide variety of applications have been considered, including fundamental
tests of quantum mechanics \cite{Vallone, Cabello}, quantum cryptography \cite{Tittel, Ribordy,
Sangouard,Berlin}, quantum communication \cite{Brendel, Marcikic1,
Marcikic2, Gisin} and clock synchronization \cite{Lamine}.
Since time and energy are continuous variables, the entanglement achieved in a specific down-conversion process is characterized by the timescales of the quantum correlations and their associated frequency bandwidths. To maximize sensitivity to time and frequency in the optical regime, it is desirable to realize broadband entanglement, where the temporal correlations can ideally be as short as the monocycle limit of a few femtoseconds \cite{Harris, Nasr, Hendrych}. However, it is difficult to characterize the temporal characteristics of entangled photons at such short timescales, since the dynamics of detectors limits the time resolution of photon detection to much longer times. Although the achievement of two-photon coherence can be confirmed by observing the temporal width of the Hong-Ou-Mandel dip for the down-converted photon pairs \cite{HOM,Giovanetti,Okamoto}, and the temporal coincidence and be confirmed by two-photon absorption, e.g. in second harmonic generation \cite{Sensarn, ODonnell}, neither method provides a complete picture of the two-photon state in frequency and time. It would therefore be desirable to have a more direct method of characterization for this specific type of entanglement.

In general, a precise characterization requires an external time-standard provided by short-time reference pulses. As we already showed in a previous work, correlated two-photon interferences with such short-time pulses can be used for a complete characterization of the two-photon state based on the four photon coincidences in the output \cite{Ren}. In the present paper, we consider an alternative procedure that makes use of the coherence between the photon pairs produced in down-conversion and the vacuum state. This coherence arises because the down-conversion process is a fully coherent non-linear process that transfers the phase of the pump beam to the two-photon coherence of the down-converted light. As a result, the two-photon wavefunction describes phase-sensitive correlations of the local single photon coherences in the two outputs. Such correlations can be measured by interfering the light in both outputs with weak coherent reference pulses and detecting the correlations between single photon outputs on both sides. Since the detection process cannot distinguish between photons originating from the references and photons originating from the down-conversion, the coincidence counts obtained in the output ports depend on quantum interferences between the two-photon wavefunction of the down-conversion and the product of the two single photon wavefunctions of the references. The spectral dependence of these interferences provides a complete phase sensitive map of the two-photon wavefunction and can be used to identify temporal features of the quantum state in terms of the periodicities of phase oscillations in frequency.

The rest of the paper is organized as follows. In section \ref{sec:spi}, we describe the characterization of a single photon wavefunction by interference with weak coherent light. In section \ref{sec:twocorr}, we describe the application of correlated measurements to down-converted light and show how the two-photon coherence is obtained from the coincidence rates. In section \ref{sec:entangle}, we describe the characteristic features expected in measurements of time-energy entanglement and consider simple experimental criteria for the evaluation of measurement data. Section \ref{sec:conclusions} summarizes the results and concludes the paper.

\section{Single-photon interference with a weak coherent reference}
\label{sec:spi}

It is well known that the interference between single photon wavefunctions from different weak coherent light sources corresponds directly to the classical interference between the corresponding fields. In the quantum formalism, this can be expressed by the formal equivalence between the operators of the field and the creation operators of the photons. In this section, we show in detail how this correspondence relates the single photon detection statistics between a reference field and an unknown coherent signal field to the single photon wavefunction of the one photon component in the signal field.

\vspace{-10 pt}
\begin{figure}
[ht]
\begin{center}
\includegraphics[width=0.3\textwidth]{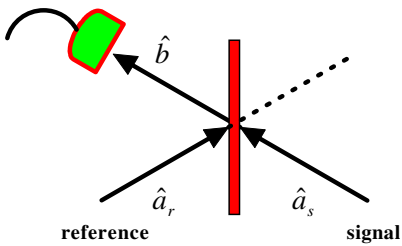}
\vspace{-5 pt}\caption{\label{fig1} Illustration of the measurement of single photon interference between the signal field and the reference field. Interference occurs because the origin of the single photon observed in the detector cannot be identified.}%
\end{center}
\end{figure}
\vspace{-12 pt}

We consider a weak optical field characterized by an unknown but
well-defined coherence in time. In the limit of photon number
averages much lower than one, the quantum state of this field can be
described by a linear superposition of the vacuum with a much
smaller single photon component, where the complete optical
coherence is described by the single photon state
$\mid\Psi_{\mathrm{s}}\rangle$. In terms of the frequency modes
$\hat{a}_{\mathrm{s}}(\omega)$, the coherent state of the signal
field can be written as
\begin{equation}
\label{eq:signal}
\mid  \Phi_{\mathrm{sig.}}\rangle=\mid
\mathrm{vac.}\rangle
+\gamma \int \langle \omega \mid \Psi_{\mathrm{s}} \rangle
 \; \hat{a}_{\mathrm{s}}^{\dag}(\omega)\mathrm{d}\omega \mid \mathrm{vac.}\rangle,
\end{equation}
where $\langle \omega \mid \Psi_{\mathrm{s}} \rangle$ is the
frequency representation of the normalized single photon
wavefunction and $\gamma$ is the overall single photon amplitude. To
characterize the unknown temporal and spectral features of the
single photon state, we use interference between the unknown signal
field and a broadband reference pulse characterized by a known
spectral decomposition $\phi(\omega)$ and an overall single photon
amplitude of $\alpha$. Here, the phases of $\phi(\omega)$ are
defined so that the peak time of the pulse is at $t=0$. Different
peak times $t_r$ can be described by multiplying each frequency
component with the appropriate phase factor. The quantum state of
this reference light can be expressed in terms of the frequency
modes  $\hat{a}_{\mathrm{r}}(\omega)$ of the reference field,
\begin{equation}
\label{eq:reference}
\mid  \Phi_{\mathrm{ref.}}(t_r)\; \rangle= \mid \mathrm{vac.}\rangle
+\alpha \int \phi(\omega) e^{i \omega t_r}
\; \hat{a}_{\mathrm{r}}^{\dag}(\omega)\mathrm{d}\omega \mid \mathrm{vac.}\rangle.
\end{equation}
The interference between signal and reference can be realized using a beam splitter as illustrated in Fig.\ref{fig1}. After the interference, photons are detected in only one of the two output ports of the beam splitter. If the measurement is spectrally resolved, the photons are detected in frequency modes $\hat{b}(\omega)$ that can be described by equal superpositions of the input modes,
\begin{equation}
\label{eq:trans}
\hat{b}(\omega)=\frac{1}{\sqrt{2}}(\hat{a}_{{\mathrm{r}}}(\omega)+\hat{a}_{{\mathrm{s}}}(\omega)).
\end{equation}
The count rate in the detector thus depends on the relative phases of the single photon components of signal and reference. For a reference peak time of $t_r$, the frequency dependent single photon count rate of the detector is given by
\begin{equation}\label{eq:power}
C_1(\omega;t_r)=\frac{1}{2}(|\alpha|^2|\phi(\omega)|^2 + |\gamma|^2|
|\langle \omega \mid \Psi_{\mathrm{s}} \rangle|^2 + 2 \alpha \gamma
\mathrm{Re}[\phi^{*}(\omega)e^{i\omega t_r} \langle \omega \mid
\Psi_{\mathrm{s}}\rangle]).
\end{equation}
In general, each frequency component $\langle \omega \mid
\Psi_{\mathrm{s}} \rangle$ of the single photon wavefunction could
be determined independently by scanning the peak time $t_r$ and
observing the corresponding interference fringes in the count rate
$C_1(\omega;t_r)$. Significantly, the peak time of the pulse defines
a phase standard that allows a reliable determination of the
coherence between different frequency components, even if the data
is obtained separately.

In many cases, it will be possible to chose a peak time $t_r$ so
different from the time of the signal pulse that the interference
term in Eq.(\ref{eq:power}) can be observed directly in the
frequency dependence of the count rate, as an oscillation with a
period of $2 \pi/(t_r-t)$ from maximum to maximum, where $t$ is the
peak time of the corresponding frequency component of the signal
pulse. If the period of oscillation varies slowly, its frequency
dependence can be interpreted as a gradual phase shift of the
amplitudes $\langle \omega \mid \Psi_{\mathrm{s}} \rangle$,
resulting in a complete determination of amplitude and phase from
interference with a single reference pulse of peak time $t_r$.
Specifically, the frequency difference $\Delta \omega$ between two
neighboring maxima determines the phase gradient of the
wavefunction according to
\begin{equation}\label{phasegrad}
\frac{d}{d\omega} \mathrm{Arg}(\langle\omega\mid\phi\rangle) = \frac{2\pi}{\Delta \omega}-t_r.
\end{equation}
From this analysis, it is easy to see that the value of $2\pi/\Delta \omega$ can be interpreted as the approximate time difference between signal and reference at a specific frequency $\omega$. In particular, it is straightforward to identify a linear chirp in terms of the correlation between time and frequency obtained from a change of $2\pi/\Delta \omega$ with frequency.

Rapid oscillations in the frequency dependence of the count rate caused by large values of the reference time shift $t_r$ in Eq.(\ref{eq:power}) also make it easier to evaluate the absolute value of the single photon amplitude $\langle \omega \mid \Psi_{\mathrm{s}} \rangle$. If the frequency difference between maximal values $C_{\mathrm{max}}(\omega)$ and minimal values $C_{\mathrm{min}}(\omega)$ is small enough, the measurement data can provide a value for the difference at each frequency, resulting in an experimentally determined value of
\begin{equation}
\label{eq:absvalue} |\langle \omega \mid \Psi_{\mathrm{s}} \rangle|
= \frac{C_{\mathrm{max}}(\omega)-C_{\mathrm{min}}(\omega)}{2
\mid\alpha \gamma \phi(\omega)\mid}.
\end{equation}
With these two equations, the single photon wavefunction of a weak coherent signal can be determined from the single photon count rates determined in interference with a known reference signal. The only condition is that the reference pulse has sufficiently high frequency components $\phi(\omega)$ throughout the bandwidth of interest.

In the single photon case, the present method corresponds to the
determination of an unknown pulse shape by interference with a known
reference pulse, where the count rate $C_1(\omega)$ corresponds to
the power spectrum of classical light. The only difference is given
by the measurement statistics: $C_1(\omega)$ is a single photon
count rate and only one value of $\omega$ can be detected in any
single run of the experiment. This difference becomes significant
when correlations between two single photon interference
measurements are considered. Coincidence counts between photons
observed at separate locations then depend on the coherence between
vacuum and two-photon states that is characteristic of the entangled
fields generated in parametric down-conversion. In the following, we
explain how correlated single photon interferences can be used to
directly determine the coherent two-photon wavefunction of the
entangled photon pairs generated by a down-conversion source.

\section{Characterizing a two-photon wavefunction by correlated single photon
interference}
\label{sec:twocorr}

The wavefunction of two-photon entangled states has no classical wave analogy, since classical waves can always be described in terms of their local amplitudes. It is therefore particularly interesting that the characterization of single photon coherence by interference with a weak coherent reference can also be applied to down-converted photon pairs, simply by taking into account the correlations between the local photon detections observed as a result of single photon interferences with two separate references. Effectively, correlated single photon interferences illustrate how the non-classical features of entanglement emerge from the application of single particle statistics to coherent phenomena.

Specifically, the down-converted light field is characterized by a
non-classical two-photon coherence that can be expressed by a
superposition of vacuum with a two-photon wavefunction. For an
overall pair amplitude $\eta$, this superposition can be written as
\begin{equation}
\mid  \Phi_{\mathrm{PDC}}\rangle=\mid \mathrm{vac.} \rangle + \eta\int \!
\langle \omega_{1},\omega_{2}\mid \Psi_{1,2} \rangle \; \hat{a}_{\mathrm{s1}}^{\dag}(\omega_1)\,{a}_{\mathrm{s2}}^{\dag}(\omega_{2})
\mid \mathrm{vac.} \rangle \mathrm{d}\omega_{1}\mathrm{d}\omega_{2},
\end{equation}
where the operators $\hat{a}_{\mathrm{s_{i}}}(\omega_i)$ represent the frequency modes of signal fields $\mathrm{s_1}$ and $\mathrm{s_2}$, corresponding to two output beams of the down-conversion process, and $\langle\omega_{1},\omega_{2}\mid\Psi_{1,2}\rangle$ describes the two photon wavefunction of the down-converted photon pairs that we wish to determine from the interferences with the short-time reference pulses.

\begin{figure}
[!htbp]
\begin{center}
\includegraphics[width=0.3\textwidth]{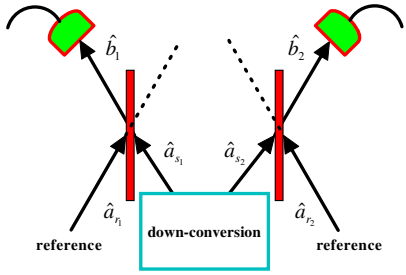}
\vspace{-10 pt}\caption{\label{fig2} Illustration of correlated measurements of single photon interference between down-converted light and two separate weak coherent references. The two photon coincidence rate is sensitive to interference, because the detection does not distinguish between photon pairs originating from the down-conversion source and photon pairs originating from the two references.}%
\end{center}
\end{figure}

As indicated in Fig. \ref{fig2}, we assume that the two output beams
are spatially separated and interfere independently with two
coherent reference pulses. In practice, the reference pulses and the
pump pulse of the down-conversion should all be derived from the
same laser source, to ensure that the phase relations between the
down-converted light and the references is fixed and constant. In
the following, we consider two reference pulses with the same
spectral decomposition $\phi(\omega)$, but different peak times
$t_{r_1}$ and $t_{r_2}$. In terms of the frequency modes
$\hat{a}_{\mathrm{r_i}}(\omega_i)$, the quantum states of the
reference fields $\mathrm{r_1}$ and $\mathrm{r_2}$ can then be
written as
\begin{equation}
\label{eq:reference} \mid  \Phi_{\mathrm{r_i}}(t_{r_i}) \rangle=
\mid \mathrm{vac.}\rangle +\alpha \int \phi(\omega_i) e^{i \omega_i
t_{r_i}} \;
 \hat{a}_{\mathrm{r_i}}^{\dag}(\omega_i)\mathrm{d}\omega_i \mid \mathrm{vac.}\rangle.
\end{equation}
After interference between the signals and the references, the equal
superpositions $\hat{b}_1$ and $\hat{b}_2$ are detected in
coincidence. Since the detection process cannot distinguish between
photons originating from the reference pulse and photons originating
from the signal pulse, the coincidence count rate depends on
interferences between the corresponding two-photon components.
Specifically, the product state of one photon from $\mathrm{r_1}$
and one photon from $\mathrm{r_2}$ interferes with the two-photon
wavefunction of the down-conversion process to provide phase
sensitive information on the two photon state. The coincidence count
rate for the frequencies $\omega_1$ and $\omega_2$ is given by
\begin{eqnarray}
\label{eq:coincidence} C_2(\omega_{1},\omega_2;t_{r_1},t_{r_2}) &=&
\mid \alpha^2
\phi(\omega_1)\phi(\omega_2)+\eta\langle\omega_1;\omega_2\mid\Psi_{1,2}\rangle\mid^2
\nonumber \\
&=& \frac{1}{4}\Big(|\alpha|^4 |\phi(\omega_1)\phi(\omega_2)|^2 +
|\eta|^2 |\langle \omega_1;\omega_2 \mid\Psi_{1,2}\rangle |^2
\nonumber \\ && +2 \mathrm{Re}\left[ (\alpha^*)^2 \eta \;\;
\phi^{*}(\omega_1)\phi^{*}(\omega_2) e^{i(\omega_1 t_{r_1}+\omega_2
t_{r_2})}\langle\omega_1;\omega_2\mid\Psi_{1,2}\rangle\right]\Big)
\end{eqnarray}
Coincidences between photon detection in output $\hat{b}_1$ and in output $\hat{b}_2$ are thus sensitive to the complete two-photon wavefunction of the down-converted light.

In principle, the complete two-photon state $\mid\Psi_{1,2}\rangle$
can be determined by scanning the peak times $t_{r_1}$ and $t_{r_2}$
to find the amplitude and the phase of
$\langle\omega_1;\omega_2\mid\Psi_{1,2}\rangle$ within the bandwidth
of the reference pulses. However, most of the relevant information
can also be obtained by using sufficiently large time differences
between the reference pulses to identify temporal information in the
interference fringes observed directly in the frequency dependence
of the two-photon coincidences. In the following, we therefore take
a closer look at the essential features of time-energy entanglement
in the coincidence data of correlated single photon interferences
and show how the temporal correlations can be determined efficiently
from the two-photon interference patterns observed in frequency.

\section{Time-energy entanglement in the experimental data}
\label{sec:entangle}

One of the reasons why time-energy entanglement between photons is
of great interest is that it occurs naturally as a consequence of
the down-conversion process. When a single photon from the pump is
converted into a pair of photons, energy conservation implies that
the pump frequency determines the sum of the down-converted
frequencies, $\omega_1+\omega_2$. Meanwhile, the frequency
difference is only constrained by spatial interferences that can be
controlled and manipulated. Broadband entanglement can be achieved
by satisfying the spatial phase matching conditions over a wide
range of frequencies. The result is a down-converted two-photon
state that can be approximately described by a narrow band
wavefunction $\psi_+$ for the sum frequency $\omega_1+\omega_2$, and
a broadband wavefunction $\psi_-$ for the frequency difference
$\omega_1-\omega_2$,
\begin{equation}
\label{eq:PDCseparate}
\langle\omega_1;\omega_2\mid\Psi_{1,2}\rangle
\thickapprox\psi_{+}(\omega_{1}+\omega_{2})\psi_{-}(\omega_{1}-\omega_{2}).
\end{equation}
For Fourier limited wavefunctions, the average arrival time of the photons has an uncertainty corresponding to the length of the pump pulse, but the time difference has a much smaller uncertainty given by the inverse bandwidth of $\psi_-$. Time-energy entanglement is therefore characterized by well-defined sum frequencies and well-defined arrival time differences for the photon pairs, in excess of the time-energy uncertainty limit for separable states given by \cite{Duan},
\begin{equation}\label{eq:limit}
\delta(\omega_1+\omega_2)\delta(t_1-t_2) \geq 1.
\end{equation}
For Fourier limited wavefunctions $\psi_-$, this simply means that
the bandwidth in $\omega_1-\omega_2$ must exceed the bandwidth of
$\psi_+$ in $\omega_1+\omega_2$, a condition that is rather easy to
satisfy in parametric down-conversion. However, the two-photon
wavefunctions obtained in down-conversion are not usually Fourier
limited, due to the phase dispersion caused by the phase matching
conditions and by the propagation through dispersive media. In a
characterization of down-converted light, it is therefore
particularly important to account for the phase differences between
the frequency components of
$\langle\omega_1;\omega_2\mid\Psi_{1,2}\rangle$.

\vspace{-10 pt}
\begin{figure}
[!htbp]
\begin{center}
\includegraphics[width=0.8\textwidth]{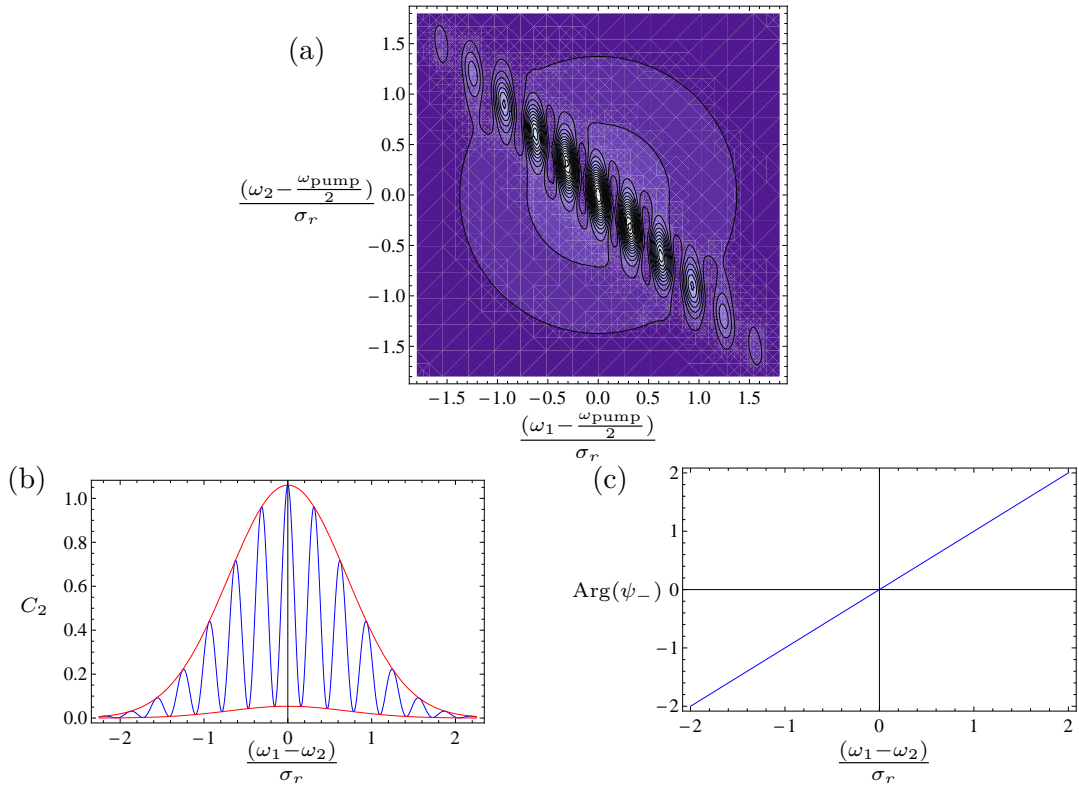}
\vspace{-10 pt}\caption{Illustration of coincidence counts obtained
for Fourier limited down-converted light with uncertainties of
$\delta(\omega_1+\omega_2)=0.2\sigma_r$ and
$\delta(\omega_1-\omega_2)=2\sigma_r$. The reference pulses have a
delay time difference of $t_{r_1}-t_{r_2}=10/\sigma_r$, centered
around the peak time of the pump pulse used in the down-conversion
($t_{r_1}+t_{r_2}=0$). (a) shows the contour plot, (b) shows the
coincidence counts for $\omega_1+\omega_2=\omega_{\mathrm{pump}}$
and the envelop functions indicating the maximal and minimal values
of the interference fringes, and (c) shows a plot of the phase
dependence of $\psi_-$ that can be determined from the periodicity
of the interference pattern in (b).}
\label{fig3} 
\end{center}
\end{figure}
\vspace{-12 pt}

 As discussed in section \ref{sec:spi}, a single
selection of reference pulse times can be sufficient to obtain the
full information about the amplitude and the phase of the photon
wavefunction if the time difference between signal and reference is
sufficiently large. In the case of down-converted photon pairs, the
arrival times of the photons are nearly equal, so a large time
difference between the reference pulses usually guarantees the
observation of rapid phase oscillations in the coincidence count
rates observed at frequencies of $\omega_1$ and $\omega_2$. Fig.
\ref{fig3} shows an illustration of the coincidence counts observed
with Gaussian reference pulses for a Fourier limited Gaussian two
photon state with a reference times difference of
$20/\delta(\omega_1-\omega_2)$, where $\delta(\omega_1-\omega_2)$ is
the uncertainty of the frequency difference distribution
$|\psi_-(\omega_1-\omega_2)|^2$. The coincidence time of the photons
can be reconstructed from the difference $\Delta_{-}$ between the
values of $(\omega_1-\omega_2)$ of two neighboring maxima of the
correlated interference pattern observed in the coincidence count
rate $C_2(\omega_{1},\omega_2;t_{r1},t_{r2})$. Specifically, the
phase change along the $\omega_1-\omega_2$ direction is
\begin{equation}
\label{eq:deltaphase} \frac{d}{d(\omega_1-\omega_2)}
\mathrm{Arg}(\psi_-(\omega_1-\omega_2)) =
\frac{2\pi}{\Delta_{-}}-\frac{t_{r_1}-t_{r_2}}{2}.
\end{equation}
If it can be confirmed that the phase depends only linearly on
$\omega_1-\omega_2$, the time correlation is given by the Fourier
limit of $\delta(t_1-t_2)=1/\delta(\omega_1-\omega_2)$. The
necessary information about the sum frequency uncertainty
$\delta(\omega_1+\omega_2)$ is provided by the width of the
interference pattern in the $\omega_1+\omega_2$ direction. It is
therefore possible to confirm that the down-converted photons
violate the uncertainty limit of separable states given by
Eq.(\ref{eq:limit}) by a factor equal to the ratio of
$\delta(\omega_1-\omega_2)$ and $\delta(\omega_1+\omega_2)$.

\vspace{-10 pt}
\begin{figure}
[!htbp]
\begin{center}
\includegraphics[width=0.8\textwidth]{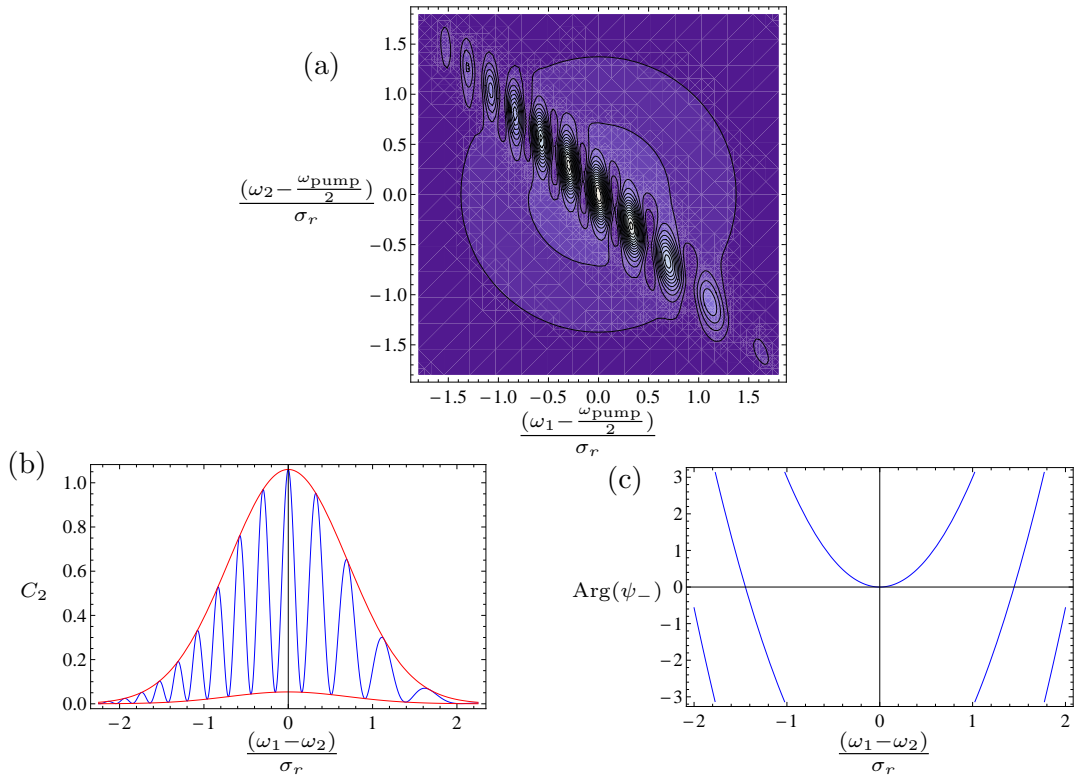}
\vspace{-10 pt}\caption{Illustration of dispersion effects observed
in the correlated single photon interferences. Parameters are the
same as in Fig. \ref{fig3}, but a phase dispersion had been added,
so that $\delta(t_1-t_2)$ is now equal to
$1/\delta(\omega_1+\omega_2)$ and the correlations of arrival times
and photon energies do not exceed the uncertainty limit of separable
states. (a) shows the contour plot, (b) shows the coincidence counts
for $\omega_1+\omega_2=\omega_{\mathrm{pump}}$ and the envelop
functions indicating the maximal and minimal values of the
interference fringes, and (c) shows a plot of the phase dependence
of $\psi_-$ that can be determined from the periodicity of the
interference pattern in (b).}
\label{fig4} 
\end{center}
\end{figure}
\vspace{-12 pt}

Any dispersion in the two-photon wavefunction will show up as a
frequency dependence of the periodicity in the interference pattern
along $\omega_1-\omega_2$. In the case of normal dispersion, the
group velocity is lower for high frequency photons and the arrival
times of high frequency photons will be delayed relative to the
arrival times of the low frequency photons. For $\omega_1>\omega_2$,
this results in positive $t_1-t_2$, and therefore in longer periods
of the interference patterns observed for positive
$t_{r_1}-t_{r_2}$. Oppositely, the periods will be shortened for
$\omega_2>\omega_1$. An illustration of the effect is shown in
Fig.\ref{fig4}.

If the dispersion effect can be expressed by a linear dependence of
the arrival time difference $t_1-t_2$ on the frequency difference
$\omega_1-\omega_2$, the uncertainty $\delta(t_1-t_2)$ can be
determined from the uncertainty of $\omega_1-\omega_2$ by
multiplication with the factor describing this proportionality. In
terms of the phase $\mathrm{Arg}(\psi_-)$ along $\omega_1-\omega_2$,
\begin{equation}
\label{eq:dispnoise} \delta(t_1-t_2) = 2 \delta(\omega_1-\omega_2)
\frac{d^2}{d(\omega_1-\omega_2)^2}
\mathrm{Arg}(\psi_-(\omega_1-\omega_2)).
\end{equation}
This equation expresses the degradation of temporal correlations
between down-converted photon pairs by a quadratic phase dispersion.
If this dispersion is very large, the correlation times may be
longer than those of Fourier limited pulses with the same sum
frequency uncertainty $\delta(\omega_1+\omega_2)$. Using the
uncertainty relation given by Eq.(\ref{eq:limit}) and the relation
given by Eq.(\ref{eq:dispnoise}), the condition for time and
frequency correlations exceeding the limit for separable photon
states can be given in terms of the phase dispersion observed in the
correlated single photon interferences. The condition reads
\begin{equation}
\frac{d^2}{d(\omega_1-\omega_2)^2}
\mathrm{Arg}(\psi_-(\omega_1-\omega_2)) < \frac{1}{2
\delta(\omega_1+\omega_2)\delta(\omega_1-\omega_2)}.
\end{equation}
The parameters in Fig. \ref{fig4} have been chosen so that they lie exactly on the boundary. The temporal correlations predicted from such data could therefore be obtained without any entanglement, e.g. from a Fourier limited Gaussian with $\delta(\omega_1-\omega_2)=\delta(\omega_1+\omega_2)$ centered at $\omega_1=\omega_2=\omega_{\mathrm{pump}}/2$.

\section{Conclusions}
\label{sec:conclusions}

We have investigated the possibility of characterizing the energy-time entanglement of
the two-photon wavefunction of photon pairs generated by parametric down conversion
using correlated single photon interference. By phase-locking the pump light with the
reference light, phase sensitive two-photon interferences can be observed in the correlated
single photon counts observed in the separate output beams. Significantly, it is possible to
obtain the complete phase information of the non-classical two-photon wavefunction from the
correlations between the coincidence counts, even though this wavefunction cannot be explained
in terms of local fields.

We have also shown how the temporal statistics of the photons can be obtained without changing the delay time of the reference pulses, by evaluating the interference patterns in the frequency spectrum obtained with sufficiently large time differences between the reference peaks and the signal photons. It is then possible to predict correlations between the arrival times of the down-converted photons without any time resolved photon detection, simply from the interferometric phases observed between the spectral components of the two-photon wavefunction and the references.

Although the method requires phase locking between the pump and the reference signals, it may be much easier to achieve high count rates than the previous proposal based on photon bunching \cite{Ren}, since it only requires two-photon coincidences. Moreover, the result can be interpreted directly in the frequency basis, with the phase information obtained from the phase of interference fringes observed either as a function of delay time, or as a rapid oscillation in the frequency dependence of the output. In this sense, correlated single photon interference gives a more direct access to two-photon coherences than alternative methods of time-sensitive quantum measurements, highlighting the possible wave aspects of photon entanglement and illustrating the role of optical coherence and dispersion effects in the non-classical statistics of entangled photon pairs. Correlated single photon interferences thus provides a potentially useful tool for the investigation of non-classical light field statistics associated with time-energy entangled photon pairs and may play an important role in the development and characterization of broad band entanglement sources and their application to optical quantum information technologies.

\section*{Acknowledgments}
This work was supported by JSPS KAKENHI Grant Number 24540427.


\end{document}